# Shock-darkening in ordinary chondrites: determination of the pressure-temperature conditions by shock physics mesoscale modeling

J. Moreau[1], T. Kohout[1,2] and K. Wünnemann[3]

1. Department of Physics, University of Helsinki, Finland (juulia.moreau@helsinki.fi)
2. Institute of Geology, The Czech Academy of Sciences, Prague, Czech Republic
3. Museum für Naturkunde, Leibniz Institute for Evolution and Biodiversity Science, Berlin, Germany.

**Abstract**

We determined the shock-darkening pressure range in ordinary chondrites using the iSALE shock physics code. We simulated planar shock waves on a mesoscale in a sample layer at different nominal pressures. Iron and troilite grains were resolved in a porous olivine matrix in the sample layer. We used equations of state (Tillotson EoS and ANEOS) and basic strength and thermal properties to describe the material phases. We used Lagrangian tracers to record peak shock pressures in each material unit. The post-shock temperatures (and the fractions of tracers experiencing temperatures above the melting point) for each material were estimated after the passage of the shock wave and after reflections of the shock at grain boundaries in the heterogeneous materials. The results showed that shock-darkening, associated with troilite melt and the onset of olivine melt, happened between 40 and 50 GPa – with 52 GPa being the pressure at which all tracers in the troilite material reach the melting point. We demonstrate the difficulties of shock heating in iron and also the importance of porosity. Material impedances, grain shapes and the porosity models available in the iSALE code are discussed. We also discussed possible not-shock-related triggers for iron melt.

**Introduction**

Shock-darkening, or shock-blackening, in ordinary chondrites (OCs) is the process that involves melting of iron sulfides and metals. Through melting they fill solid silicate grain boundaries and cracks to form a network of tiny melt veins. It leads to optical darkening of large volumes of OC material and it has been observed in earlier reports on OCs (Heymann 1967; Britt et al. 1989; Britt and Pieters 1989, 1994; Keil et al. 1992) and more recently in the Chelyabinsk LL5 ordinary chondrite (Kohout et al. 2014). In contrast, silicates such as olivine and pyroxenes are found to be highly shocked but not substantially molten. The few percent of silicate melt forms melt pockets. Sometimes, larger localized impact melt veins can intercalate between the shocked solid material. The exact degree of shock metamorphism (Stöffler et al. 1991) of the silicate material is difficult to evaluate under optical microscopy because the shock-darkening turns the material opaque. Our work aims to study the pressure range at which shock-darkening occurs.

This process has a significant effect on reflectance spectra of meteorites as seen, e.g, in the Chelyabinsk meteorite (Fig. 1, Kohout et al. 2014). The Chelyabinsk meteorite light lithology spectra resemble S-complex asteroids spectra while the dark lithology spectra are flatter and darker and appear similar to some C/X-complex asteroids spectra. This observation combined with asteroid spectra classification and distributions (DeMeo et al. 2009; DeMeo and Carry 2014) suggests that some C/X-complex asteroids may, indeed, be shock-darkened S-complex asteroids (Britt and Pieters 1989; Britt et al. 1989; Kohout et al. 2014). However, shock-darkening should not to be confused with darkening occurring on asteroid surfaces (space weathering, Clark et al. 2002, and implantation of solar wind ions as observed in gas-rich OCs, Britt and Pieters 1994).

Shock-darkening may be a consequence of shock compression and associated heating. Thus, we use principles of shock physics (Melosh 1989; Zel'dovich and Raizer 2002) and hydrocode modeling (Collins et al. 2013) to quantify processes related to the propagation of a shock wave in OC-like samples.

**Methods**

We modeled shock-compression in heterogeneous materials mimicking OC compositions. We studied the thermodynamic processes and assessed melt fractions for each integrated mineral phase. We determined the shock-darkening pressure range for the onset of melting of metals or iron sulfides phases with minimum melting of silicates. We used the iSALE-2D shock physics code (Wünnemann et al. 2006), which is based on the SALE hydrocode solution algorithm (Amsden et al. 1980). To simulate hypervelocity impact processes in solid materials, SALE was modified to include an elastic-plastic constitutive model, fragmentation models, various equations of state (EoS), and multiple materials (Melosh et al. 1992; Ivanov et al. 1997). More recent improvements include a modified strength model (Collins et al. 2004) and a porosity compaction model (Wünnemann et al. 2006; Collins et al. 2011).

We used a planar 2-D Eulerian frame of reference (Collins et al. 2013) to simulate on a mesoscale the propagation of planar shock waves. Analogous to so-called shock recovery experiments (e.g. Langenhorst and Deutsch 1994; Langenhorst and Hornemann 2005) we generated the shock wave by impacting a flyer plate on top of a series of layers. Within this setup we observed peak shock pressures and pressure reflections between mineral phases (sample). This approach has been previously used to describe shock propagation in heterogeneous materials by e.g. Crawford et al. (2003), Ivanov (2005), Riedel et al. (2008), Borg and Chhabildas (2011), Güldemeister et al. (2013), and Bland et al. (2014).

*Materials, model setup and Equations of State (EoS)*

Following characteristics of OCs (McSween et al. 1991; Stöffler et al. 1991; Rubin 1992, 2010; Friedrich et al. 2008; Oshtrakh et al. 2008; Kohout et al. 2014; Scott and Krot 2014; Righter et al. 2015) we selected the following components for the mesoscale setup (meso-grains model, MG):

1. Olivine: used as the rock matrix. It is the dominant silicate phase in OCs and was preferred over pyroxenes. We assumed a porosity of 6% for olivine (Britt et al. 2002; Consolmagno et al. 2008; Kohout et al. 2014; Ruzicka et al. 2015).
2. Iron: it mimicked mechanical and thermal properties of the FeNi alloy.
3. Troilite: used as the iron sulfide phase. We used the mechanical properties of pyrrhotite as a proxy for troilite.

The four layers of the MG model, composed of olivine, were (similar to the setup from Bland et al. 2014):

1. The flyer plate (600x600 cells) that impacted the sub-layers to generate the shock wave. The shock wave propagated at a velocity proportional to the generated shock pressure downwards into the buffer plate and upwards into the flyer plate. The shock plateau did not attenuate as any dissipative processes had been neglected in the model. Unloading was only caused by the rarefaction wave that had originated from the free surface of the flyer plate and propagated downwards. The vertical extent of the flyer plate had been chosen sufficiently thick so that the length of the shock plateau exceeded the thickness of the sample. This allowed reflections of the shock wave at the interfaces of different materials (iron, troilite, olivine) in the sample before unloading by the rarefaction wave occurred. We defined the constant nominal pressure, the pressure in a homogeneous sample that was only composed of the matrix material (olivine). It was used as the scale of our study and to represent our results.
2. The top buffer plate (600x360 cells) from which the nominal pressure defined above was recorded.
3. The sample plate (600x720 cells) that comprised rounded troilite and iron grains of different sizes (between 24 and 72 cells per grain diameter, CPGD), randomly distributed in the olivine matrix. Peak shock pressures in the materials were recorded by Lagrangian tracers.
4. The bottom buffer plate (600x240 cells) let the shock wave progress and preserve the shock pulse in the sample plate. It allowed to record all reflections events and peak shock pressures in the materials.

Fig. 2 shows schematically the MG model. Initial temperature of the setup was 293 K. We did not ran the models until the release state (when the sample is completely unloaded from shock pressure).

We used the Analytic EoS (ANEOS, Thompson 1990; Melosh 2007) and the Tillotson EoS (Tillotson 1962; Melosh 1989; Brundage 2013) to calculate the thermodynamic behavior of the different materials during shock compression and subsequent release. We detail below our choice for the parameters describing the thermodynamic and petrophysical properties of the different materials. In the Tillotson EoS, parameters are: $a, b, \alpha, \beta, B$ (Pa), material bulk modulus $A$ (Pa), initial and incipient vaporization material densities $\rho_0$ and $\rho_{IV}$ (g/cm$^3$), initial energy, incipient and complete vaporization material energies $E_0$, $E_{IV}$, $E_{CV}$ (J/kg). While some parameters may be considered as fitting parameters, all others have a physical meaning. Parameters are compiled in Table S2 from the on-line Supporting Information file.

*Olivine solid-solution series*

Because olivine is a solid-solution found in OCs (forsterite-fayalite, $Fo_XFa_{100-X}$), we first evaluated the two options ANEOS (fixed composition, $Fo_{90}$) and Tillotson EoS (range of composition). We varied the thermal properties separately.

We derived the Tillotson EoS parameters from $Fo_{75}$ (Marinova et al. 2011). We used linear interpolations from densities $\rho$, (Deer et al. 1997), and bulk moduli $A$ (Zha et al. 1996 and Núñez-Valdez et al. 2010). Because ANEOS data correlate well with Hugoniot literature data (Fig. 3, particle and shock wave velocities in olivine) we used ANEOS instead of the Tillotson EoS in all simulations.

We varied the heat capacity $c_p$ (J/kgK) at 20 ºC according to the percentage in forsterite (Waples and Waples 2004). We interpolated the range in melting temperatures ($T_{melt}$) at zero pressure (2162 K for forsterite and 1487 K for fayalite, Weizman et al. 1997) linearly to the solidus liquidus binary phase intermediate as a percentage of forsterite.

We did not use pyroxenes. Hugoniot data for pyroxene $En_{85}Fs_{15}$ (Trunin 2001) are close to the ANEOS calculated Hugoniot data for olivine $Fo_{90}$ as seen in Fig. 3. The melting temperature for enstatite is 1830-1832 K (Boyd et al. 1964; Presnall and Gasparik 1990).

*Iron sulfides and metals*

Troilite and iron are the main phases generating shock-darkening. We used ANEOS for iron. However we needed an EoS for troilite, for which we employed the Tillotson EoS.

We derived the Tillotson EoS parameters using available Hugoniot data for pyrrhotite (Ahrens 1979, Brown et al. 1984), similar to troilite. We adapted the density (pyrrhotite, 4611 kg/m³), the bulk modulus (45 GPa, originally 80 GPa in Martin et al. 2001), the internal energy (14.343 MJ/kg, Martin et al. 2001) and other parameters to fit the Tillotson EoS data to the available Hugoniot data. Fig. 4 shows the Tillotson EoS Hugoniot curves compared to the Hugoniot literature data.

The specific heat capacity of iron is 449 J/kgK (Lide 2003) and similar to kamacite FeNi alloy found in meteorites. We set the specific heat capacity for troilite to 619.23 J/kgK (average value from LL-OC, Mare et al. 2014). The melting temperature of iron at zero pressure is 1825 K (Zhang et al. 2015, similar to kamacite, Cacciamani et al. 2006; Garrick-Bethell and Weiss 2010) and troilite 1463 K (Lide 2003). These melting temperatures are higher than the metal and iron sulfide mixture eutectic point (Tomkins 2009; Mare et al. 2014). However, in un-shocked material, mixtures of metals and iron sulfides are scarce and metal or iron sulfide grains are mostly found isolated or in single contact zones. These mixtures are mostly created after melting where the molten iron sulfides and metals can mix together (Schmitt 1995, 2000). In our study, we determined a precursor material with no such mixture.

*Strength properties*

We used the von Mises yield strength criterion for olivine, where the shear strength *Y* (1.5 GPa, Brace and Kohlstedt 1980) is a constant and does not depend on pressure or strain rate. For iron and troilite, we assumed hydrodynamic behavior and neglected any effects of strength. Although the very simple strength model of olivine and the hydrodynamic behavior of iron and troilite appear to be a relatively crude approximation, we have found in test runs that the effect of strength was negligible for the determination of shock pressures. Because we did not run the models to the release state, the recorded peak shock pressures during compression were very little affected by strength. We did not consider thermal softening in our models.

*Porosity*

Porosity can be considered in two different ways in our simulations depending on the scale of porosity:

1. On a mesoscale (meso-pores model, MP), individual pores of different shapes and sizes are resolved (Kowitz el al. 2013; Güldemeister et al. 2013) in a setup similar to MG models without additional particles.

2. On a microscale, when pores are significantly smaller than grains and cannot be resolved in the same model, porosity is treated as a state parameter. Thus the olivine EoS was combined with the ε-α compaction model (Wünnemann et al. 2006). The input parameters for olivine are shown in Table 1 (Bland et al. 2014) and their physical meaning is explained in detail in Collins et al. (2011).

*Post-shock temperatures and melt assessment*

As mentioned above we did not run the models until the sample was completely unloaded from shock pressure to record post-shock temperatures (release state method). To estimate post-shock temperatures we utilized the simple relationship between peak shock pressures (peak shock pressure method) and post-shock temperatures (as described e.g. in Artemieva and Ivanov 2004, and Fritz et al. 2005). We describe in details this method with a comparison to the release state method in the on-line Supporting Information file.

The linear relationship between particle velocity, $u_p$ (m/s), and shock wave velocity, $U$ (m/s), used in the peak shock pressure method, is the basis to compute post-shock temperatures in materials. This relationship requires two parameters, C (m/s) and S, to define a line in the $u_p$-$U$ space:

$U = C + S \cdot u_p$  (1)

These parameters have been determined from linear regressions using Hugoniot data generated by iSALE. Material specific *C* and *S* parameters used in the study are compiled in Table 2. Because of solid state phase transformations, iron and olivine have been approximated by two line segments in $u_p$-$U$ space. Fig. 5 shows the resulting post-shock temperatures in the studied materials.

In our study we limited post-shock temperature assessment to a constant heat capacity (required in the calculations). However, heat capacity varies with temperature (forsterite, Gillet et al. 1991 – iron, Desai 1986) and the energy required to heat the material may be higher. For troilite (Fig. 5, Chase 1998) we observed that the required pressure to reach the melting temperature is approximately less than 10% higher than the required pressure using constant heat capacity. This error is still acceptable within the general uncertainties intrinsic for the hydrocode modeling approach.

In a final step, we compared the post-shock temperature to the melting temperature $T_{melt}$ at normal pressure in each tracer. The ratio between the number of tracers reaching temperatures $\geq T_{melt}$ and total number of tracers (Wünnemann et al. 2008) was used as a first-order approximation to estimate a melt fraction in % in a given area. We did not include heat of fusion to account for partial melting and temperatures were overestimated over the melting temperature.

*Resolution*

We determined the required resolution in the MG models using a setup with 4% of randomly distributed troilite grains ranging in size from 40 to 120 μm in the olivine matrix. We assumed strengthless and non-porous material. The resolution parameter was the number of cells per grain diameter (CPGD). In a series of simulations we varied the CPGD number between 8 and 48 and obtained the percentage of tracers with post-shock temperatures $\geq T_{melt}$ in troilite. The nominal pressure was sufficiently high to generate melt fractions between 5 and 95%. We ran each scenario multiple times to account for stochastic effects (random distribution of troilite grains).

We obtained less variation in the melt fraction at higher resolutions. Detecting <1% olivine tracers with temperatures $\geq T_{melt}$ was also a means to assess if the resolution was sufficient to detect small amount of melt fraction (occurring >36 CPGD). The results are shown in Table 3 and in Fig. 6. We considered a resolution of 48 CPGD to be adequate in all the models. Layers and grains sizes are stated in Fig. 2. The size of a cell for the chosen resolution was 1.6 μm.

*OC material approximation*

We represented the different OC's, H, L and LL types, by varying olivine thermal properties as a function of fraction in fayalite (Righter et al. 2015):
1. H: olivine Fo$_{82}$; melting point, 2049 K; heat capacity, 814 J/kgK.
2. L: olivine Fo$_{76.5}$; melting point, 2011 K; heat capacity, 816 J/kgK.
3. LL: olivine Fo$_{71}$; melting point, 1973 K; heat capacity, 817 J/kgK.

We used data from McSween et al. (1991) to determine the distribution of the metals and iron sulfides in each OC type and expressed them as percentages of iron and troilite. The typical distribution of iron and troilite in the OCs is shown in Fig. 7 where geometric means of the ratios are used (Crawley 2005). We emphasized here the strong variability in iron fraction.

**Results**

Using the MG model we generate two sets of results for each OC's type, H, L and LL, at different nominal pressures:
1. Assuming a porous olivine matrix with the iron and troilite grains (referred as MG model).
2. Assuming a non-porous olivine matrix in the MG model which we combine with results of MP models (referred as MG/MP model). The MP models provide additional information on the generation of additional heat in olivine by crushing of individual pores. In these models we randomly distributed pores in the sample plate, where each pore was ~6, 10 or 16 cells in size, respectively (corresponding to ~10, 16 and 26.6 μm). The total amount of pore space was 6%. The post-shock temperatures estimation for the different pore sizes are shown in Fig. 8. The relatively large standard deviations are a result of strong localized heating. Similar hotspots have been observed in mesoscale modeling of shock wave propagation in porous sandstone by Güldemeister et al. (2013).

Figs. 9 and 10 show examples of the MG model simulations in porous olivine. The figures depict snapshots of peak shock pressures and areas of tracers reaching $T_{melt}$ for each OC type at two different nominal pressures. Figs. 11 (MG model, 8 different scenarios) and 12 (MG/MP model, 7/5 different scenarios) show results of the simulations as a function of nominal pressure for each OC – each scenario was run 3 times (random distribution of grains). The different lines in Fig. 11a show the fractions of tracers reaching $T_{melt}$ for each phase. Fig. 11b combines the fractions of tracers reaching $T_{melt}$ for iron and troilite phases. In Fig. 11c the average peak shock pressures in olivine are plotted for different iron fractions in olivine. In Fig 12, the red dashed lines are results from the MP models only. Detailed results can be found in the Supporting Information File (tables S3-S8) in which peak shock pressures are averages for all tracers in each material.

All models show the following results in terms of shock melting:

1. In troilite, at 37 GPa nominal pressure (~45 GPa peak shock pressures in troilite), a small fraction of tracers start to reach $T_{melt}$ until all tracers reach $T_{melt}$ at a nominal pressure of 52 GPa (corresponds to ~70 GPa peak shock pressures in troilite). These results are a consequence of an increase in entropy by single shocks in most cases. Reflections from iron only slightly influence the peak shock pressures in troilite (as seen for olivine in Fig. 11c).

2. In olivine, fractions of tracers start to reach $T_{melt}$ at ~50 GPa (nominal pressure). However, this is a consequence of reflected shock waves at iron grain boundaries in MG model (as a result of high impedance contrast between olivine and iron) that ramps up the nominal pressure by ~10 GPa (H-OC). 10% to 60% of tracers reach $T_{melt}$ at 63 GPa nominal pressure, which is proportional to the iron grain abundance in the LL-, L- and H-OC models (Fig. 11c). Because olivine is non-porous in the MG/MP model, it requires higher pressures for heating (see Fig. 8). However, the crushing of pores and, thus, the generation of additional heat, results in temperatures > than $T_{melt}$ in olivine tracers for nominal pressures of ~50 GPa (Fig. 12), as observed in the original MG model.

3. Only localized clusters of tracers reach $T_{melt}$ in iron between 50 and 58 GPa (nominal pressure). It is due to strong reflections from proximal iron grains causing a significant increase in peak shock pressures. Thus, it depends on the iron grains distribution in the sample plate. The number of tracers reaching $T_{melt}$ remains, however, small (<10% at ~65 GPa in the LL-OCs). It is never the case that all tracers reach $T_{melt}$ in iron in any model (the maximum nominal pressure in the suite of numerical experiments for this study was 63 GPa in MG model and 67 GPa in MG/MP model).

Tracers in troilite start to reach $T_{melt}$ from ~40 GPa until they all reach $T_{melt}$ at ~52 GPa. With tracers in olivine starting to reach $T_{melt}$ at ~50 GPa (in MG and MG/MP models), we set the upper limit for shock-darkening at 50 GPa - within a positive error of ~10% regarding the post-shock temperatures calculations for troilite as explained in the method section. We do not consider iron to cause shock-darkening in our study because few tracers in iron reach $T_{melt}$ when large fractions of tracers in olivine have already reached $T_{melt}$.

**Discussion**

The modeled shock effects, particularly those associated with post-shock melting of troilite, resemble shock-darkening observed in OCs. With the large fractions of troilite reaching $T_{melt}$, the modeling reproduced the shocked chondrite fabric responsible for the attenuation of reflected light spectra. The models results on shock metamorphism are in general agreement with the shock-metamorphism in OCs described by Stöffler et al. (1991) and also the shock recovery experiment studies described by Schmitt (1995, 2000). The fractions of tracers reaching $T_{melt}$ in olivine in our scenarios were either caused by reflections at the iron grain boundaries or by the closure of pores. The latter (MP models results) yielded fractions larger than reported by Schmitt (2000) for H-OC. This deviation is explained by uncertainties in our temperature estimates. At pressures between 30 and 60 GPa, Schmitt (2000) also reported that troilite and/or metals had either melted into small droplets, larger melt pockets or also into melt veins in the olivine cracks (shock-darkening) formed by brittle fracturing in the wake of the shock wave. Other observations of iron sulfide and metal melts have been reported at 50 GPa shock pressure in a H-OC (Xie et al. 2014).

Our study provides a more quantitative understanding of the role of iron and troilite grains in shock metamorphism. Earlier studies by Stöffler et al. (1991) and van der Bogert et al. (2003) proposed that shearing between silicate, metal and iron sulfide grains may have been the mechanism causing melting of iron sulfides and metals (frictional heating). They also mentioned the presence of opaque (dark) melt veins composed of silicates at low shock stage S3-S4 that had developed into networks of veins. However, these effects are only local. Shock-darkening related to shock metamorphism can affect large volumes of material.

It is common to find evidence for partial melting of silicates in shock-darkened material. This has been observed in our models, where fractions of olivine tracers reached $T_{melt}$ at iron grain boundaries. Also the shock-darkening in our study was linked to the large fractions of tracers reaching $T_{melt}$ in troilite, which is consistent to troilite melting upon low shock pressures (Ahrens 1979; Mang et al. 2013). However, Stöffler et al. (1991) observed that shock-darkening had also involved melting of metals in some OCs, unlike our study.

For this reason, it may be necessary to consider other mechanisms for complete or partial melting of iron phases than pure shock-induced melting (which would require higher pressures than in our study, Brown et al. 1984; Ahrens et al. 1998). Here we discuss additional mechanisms responsible for iron heating or melting, and other effects, that we did not account for in our models:

1. All grains in our models were rounded, varying in size (test results on rectangular grains showed no differences). However, troilite and iron grains in meteorites occur at arbitrary shapes: from rounded to angular, with high or low sphericity, elongation transversal or longitudinal to the shock wave, aggregated together or not. In a series of models we tested the effect of grain elongation and orientation (oblate, transversal and prolate, longitudinal to the shock wave) on shock pressure amplification by reflections (using 20% of iron grains at 45 GPa nominal pressure). Fig. 13 shows the distribution of pressures inside iron grains and in the olivine matrix. Strong reflections (as a result of impedance contrast, Hirose and

Lonngren 1985; Kinslow and Cable 1970; Ahrens 1993) occurred at the long axis of oblate iron grains and short axis of prolate grains. We observed that prolate iron grains had been highly shocked to their center as well (Fig. 13). This phenomenon was caused by uneven shock wave fronts occurring along the iron-olivine boundaries. It had generated two reflected shock waves on opposite boundaries that collided at the grain center and enhanced peak shock pressures (Davison 2008). Fig. 14 summarizes the peak shock pressures observations for all kinds of iron grains orientation/elongation in olivine matrix and iron grains. In Fig. 15 the frequency distributions of pressures inside the iron grains for the two extreme prolate and oblate elongations are shown. Although locally shocked, we observed lower average peak shock pressures in prolate grains. In case of aggregated troilite and iron grains (commonly found in meteorites, Schmitt 1995, 2000), shock pressure enhancement can happen between the interfaces as a result of impedance contrasts. However, shock pressure enhancement may be not sufficient to heat iron to its melting point. Alternatively, Güldemeister et al. (2013) have shown that the collapse of pores can generate local hotspots, where the shock pressure is amplified by up to a factor of 4. Although this is a very localized effect, adjacent iron grains may undergo strong heating in such a scenario.

2. Thermal diffusivity (heat conduction) is not considered in the iSALE code. However, it can influence the heating process of material as a consequence of thermal anomalies (hotspots, grain boundaries heating). On olivine and iron, because of higher shock entropy in olivine, heat will transfer from the hotter olivine material to the colder iron grain. As thermal diffusivity is hihgly dependent on temperature (influencing thermal conductivity and heat capacity), during the shock stage it will also be dependent on pressure (influencing density). At initial conditions, thermal diffusivity in iron is ~25 mm$^2$/s. However, at shock pressures of ~60 GPa, thermal diffusivity can be reduced to less than 5 mm$^2$/s. Despite the material in our models is under shock for <1μs, this duration is higher in impact events (e.g. roughly <0.1 s for a 500 m impactor at 6 km/s velocity or <1 s for a 2 km impactor at 6 km/s). Because iron grains are smaller than a millimetre, thermal diffusivity from olivine to iron happens before material is released from shock. As a rough approximation, a Fourier number of 1.25 is found if an iron grain has diffusivity of 5 mm$^2$/s, radius of 200 μm and the elapsed time is 0.01 s, assuming that the diffusion of heat into iron is significant during the shock stage (in impact events). A more accurate assessment of the effect of thermal diffusion whether during or after shock compression, when localized hotspots occur as a consequence of heterogeneities and presence of pore space, is beyond the scope of this paper.

3. The presence of an eutectic mixture between iron sulfides and metals (Kullerud 1970; Sharma 2004; Xie et al. 2014) is not negligible as the melting points of both materials decrease to a common value. If present in precursor OC material, it could lead to the melting of iron.

Besides the lack of some physical processes (e.g. heat conduction) in our models, and temperature estimation errors (see Methods section), the determination of the post-shock temperatures comprises some simplifications that require further discussion. The peak shock pressure method to determine post-shock temperatures assumes that the material experiences its maximum pressure (peak shock pressure) by a single shock pulse. However, the peak shock pressure may result from the superposition of several reflected shock waves, resulting in a smaller increase in entropy (and post-shock temperature) than in case of a single shock

(Langenhorst and Hornemann 2005). The complexity of the shock wave propagation in the sample plate is shown in Fig. 16 (H-OC type model), where instant pressures and recorded peak shock pressures are shown. It can be observed that the primary shock wave was reflected at iron or troilite grain boundaries. The secondary (reflected) shock wave propagated into an already pressurized and compressed area. It often happened that reflected pressures were lower than recorded peak shock pressures (orange arrows in Fig. 16a). The observed shock front irregularities were caused by velocity contrasts within grains.

Fig. 17 illustrates the complexity of these reflections in the sample plate (same scenario as in Fig. 16). The peak shock pressure plateaus (i.e. reflections) recorded by tracers are shown in Fig. 17a. The differences between first and last peak shock pressures are shown in Fig. 17b. Three profiles of recorded peak shock pressures in tracers are shown in Fig 17c. From these observations we conclude:

1. Iron is mostly instantly shocked to the final peak shock pressure.
2. Troilite shows the same characteristics as iron, but it is more dependent on the iron grain distribution (mostly in H-OCs).
3. Olivine shows several reflexions and differences between first and last recorded peak shock pressures are high (~30 GPa).

From these observations we consider the peak shock pressure method calculations, explained in the Methods and in the on-line Supporting Information, to be more accurate for iron or troilite than for olivine. We tend to overestimate heating in olivine despite the occurrence of reverberation effects.

If we account for the uncertainties on assessing temperatures and melting of the materials, detailed in the Methods section and this section, these uncertainties could be balanced by effects that are not introduced in the iSALE code such as frictional heating between grains and other thermal and mechanical effects that we discussed earlier. Finally, we want to emphasize that our iSALE modeling does not provide any information on the subsequent migration of the melt.

**Conclusions**

Our results illustrate the importance of metals and iron sulfide grains in the thermodynamics involved in shock compression of ordinary chondrites. The main agent in the OCs shock-darkening is the troilite. It easily reaches melting temperature under shock pressure between 40 and 50 GPa (nominal pressure in the sample). Olivine starts to reach melting temperatures at pressures >50 GPa and this is dependent on the iron content. We find that iron is difficult to melt by shock pressure only. We discuss that heat conduction, pore closure, grain shapes, inclusions (eutectic melt) and grain distribution in the samples are factors that may contribute to melting of metals. It also could balance the uncertainties in estimation of temperatures and the melting of the materials. Thus shock-darkening is not solely induced by shock but it can occur once the thermal conditions to melt troilite and iron are met.

*Acknowledgements* - We gratefully acknowledge the developers of iSALE-2D, including Gareth Collins, Kai Wünnemann, Dirk Elbeshausen, Boris Ivanov and Jay Melosh. The work was supported by Academy of Finland Project No. 293975. Institute of Geology, the Czech Academy of Sciences is supported by Ministry of Education, Youth and Sports project no. RVO67985831. We also thank the team of the Museum für Naturkunde in Berlin for their ideas, their good advice and the pleasant time we spent together. We finally acknowledge Alisdair McLean for his comprehensive revision of the English of the original manuscript, reviewers Boris Ivanov and Axel Wittmann for their important comments on the manuscript, Natalia Artemieva for her precious help on finalizing the manuscript and Ilmo Kukkonen for his insights on the thermal effects. Some plots in this work were created with the pySALEPlot tool written by Tom Davison.

Table 1. Porosity parameters for olivine $Fo_{90}Fa_{10}$.

| | |
|---|---|
| **Distension $\alpha = 1/(1-\Phi_{porosity})$** | 1.0638 ($\Phi = 6\%$) |
| **Compaction rate $\kappa$** | 0.94 |
| **$V_{sound}$ ratio from porous to solid material ($\chi$)** | 1 |
| **Distension to power law $\alpha_X$** | 1.02 |
| **Volume strain at plastic compaction** | $-1*10^{-5}$ |

Table 2. Compilation of *C* and *S* material parameters.

| | $C_1$ (m/s) | $S_1$ (slope) | Transition (GPa) | $C_2$ (m/s) | $S_2$ (slope) |
|---|---|---|---|---|---|
| *$Fo_{90}Fa_{10}$ (ANEOS)* | 6382.04 | 0.838 | 69.2 | 3398.113 | 1.780 |
| *$Fo_{90}Fa_{10}$ (ANEOS) (6% por.)* | 4145.2 | 1.621 | 71.15 | 4264.82 | 1.41 |
| *Iron* | 4859.20 | 1.332 | 7.35 | 3572.65 | 1.824 |
| *Pyrrhotite/Troilite* | 3226.823 | 1.4593 | - | - | - |

Table 3. Comparison of model results for various iron sulfide grain resolutions.

| Resolution | CPGD[1,2] | Tracers > $T_{melt}$ (%)[1] | Standard deviation (%) | Number of runs |
|---|---|---|---|---|
| 1st | 8 | 24.09 | 2.39 | 10 |
| 2nd | 12 | 34.35 | 1.03 | 10 |
| 3rd | 16 | 37.48 | 0.75 | 10 |
| 4th | 24 | 41.42 | 1.07 | 7 |
| 5th | 36 | 43.53 | 0.9 | 5 |
| 6th | 48 | 44.77 | 1.06 | 3 |

[1] The tests were carried out with troilite.
[2] Cells per grain diameter

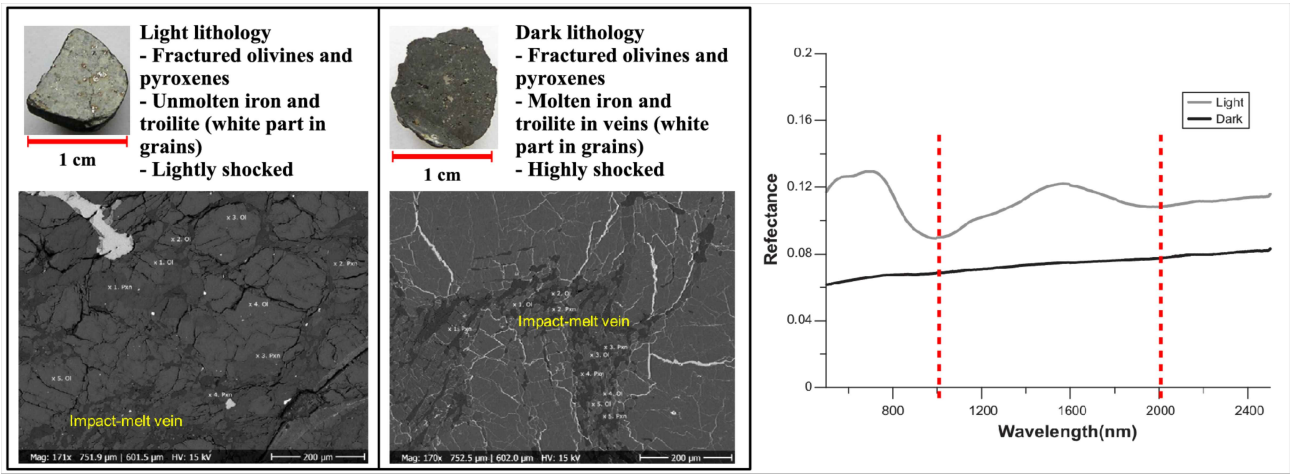

Fig. 1. Description of the light and shock-darkened lithologies of two samples taken from the Chelyabinsk LL5 meteorite and the resulting reflectance spectra. The dashed line indicates the 1 and 2 µm bands. Modified from Kohout et al. (2014).

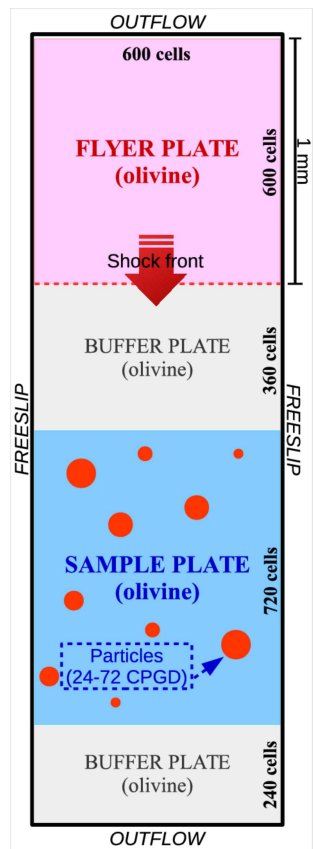

Fig. 2. Schematic of the mesoscale setup used in the study. The empty 1 cell space on top of the flyer plate allows the shock wave to be released (rarefaction wave).

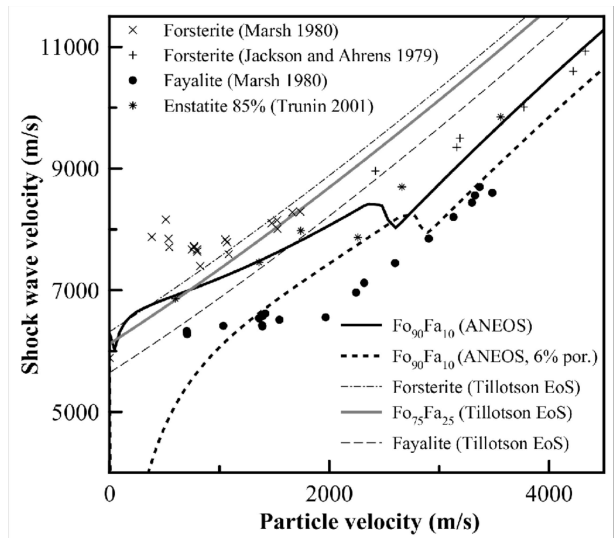

Fig. 3. Particle velocity / shock wave velocity Hugoniot data for several compositions in olivine and enstatite generated and compiled from ANEOS, Tillotson EoS and the literature. The ANEOS and literature data show phase changes. Tillotson EoS generated Hugoniot data do not represent the phase change along the profile.

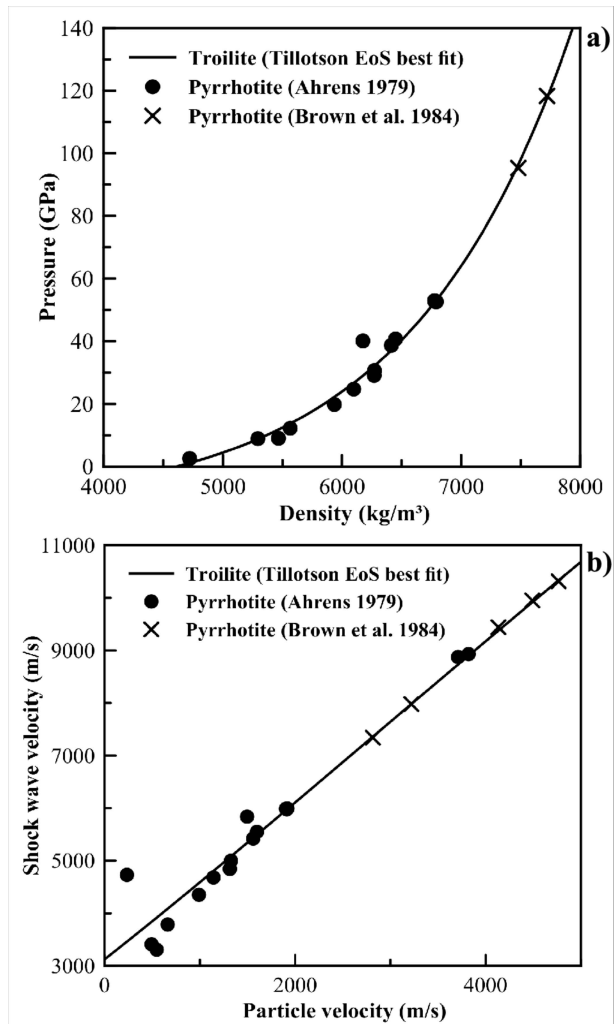

Fig. 4. Hugoniot data and Tillotson EoS Hugoniot fits for troilite using pyrrhotite data obtained from the literature with a) density and pressure and b) particle velocity and shock wave velocity.

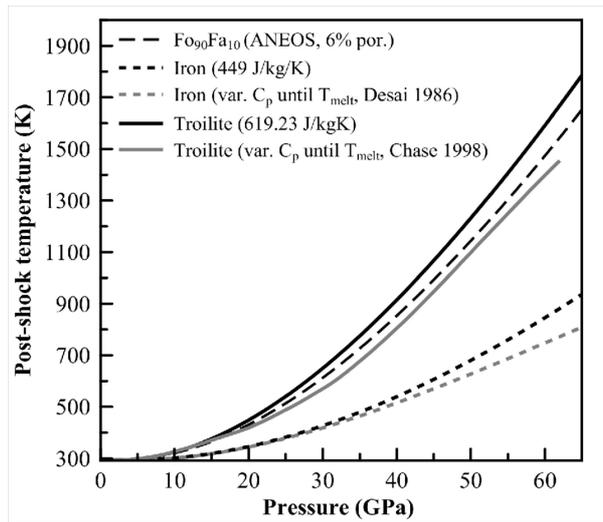

Fig. 5. Post-shock temperatures compilation of the studied materials. The additional curves for troilite and iron (grey) are examples in case varying heat capacity was considered in the post-shock temperature calculation, up to the melting point. Although not shown, the alternative curve for iron shows a strong deviation in higher pressures. Temperature is raised by 293 K to meet the initial temperature conditions and because post-shock temperatures are *ΔT*.

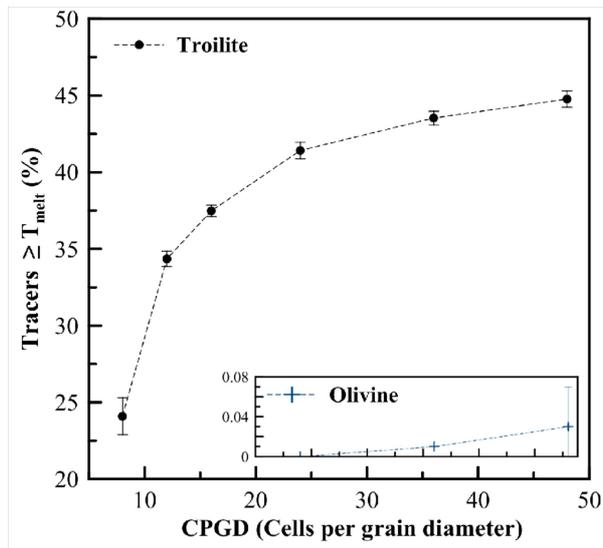

Fig. 6. Resolution tests on troilite grains. The variations in the amount of tracers reaching melting temperature at higher resolutions are reduced and results approach an asymptotic value. The inset shows values for olivine. The error bars indicate standard deviations.

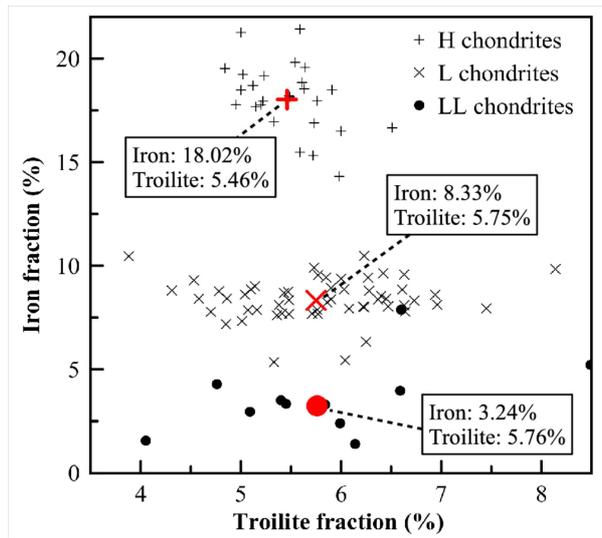

Fig. 7. Troilite and iron fractions in ordinary chondrites after McSween et al. (1991). Mean values are shown for each OC type.

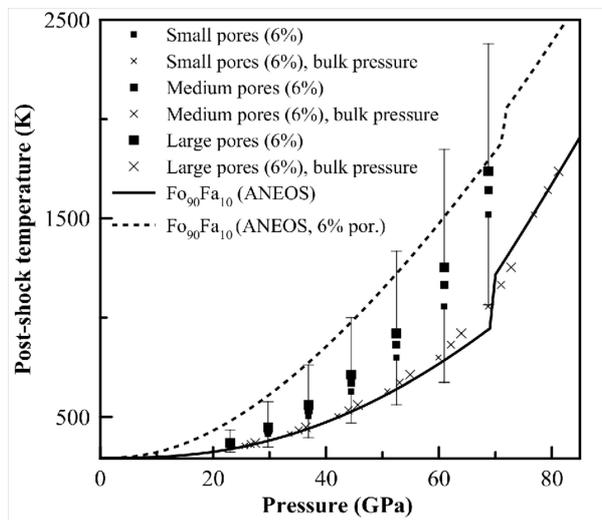

Fig. 8. Comparison between theoretical post-shock temperature calculation of porous and non-porous olivine with a MP model approach. The upper standard deviation bars are used as relative indicators of the larger pores and the lower standard deviation bars are relative indicators of the smaller pores. The crosses depict the peak shock pressures. They are located close to, the non-porous olivine ANEOS used in the meso-pores setup. Thermal properties of H-chondrite olivine have been used for the calculations.

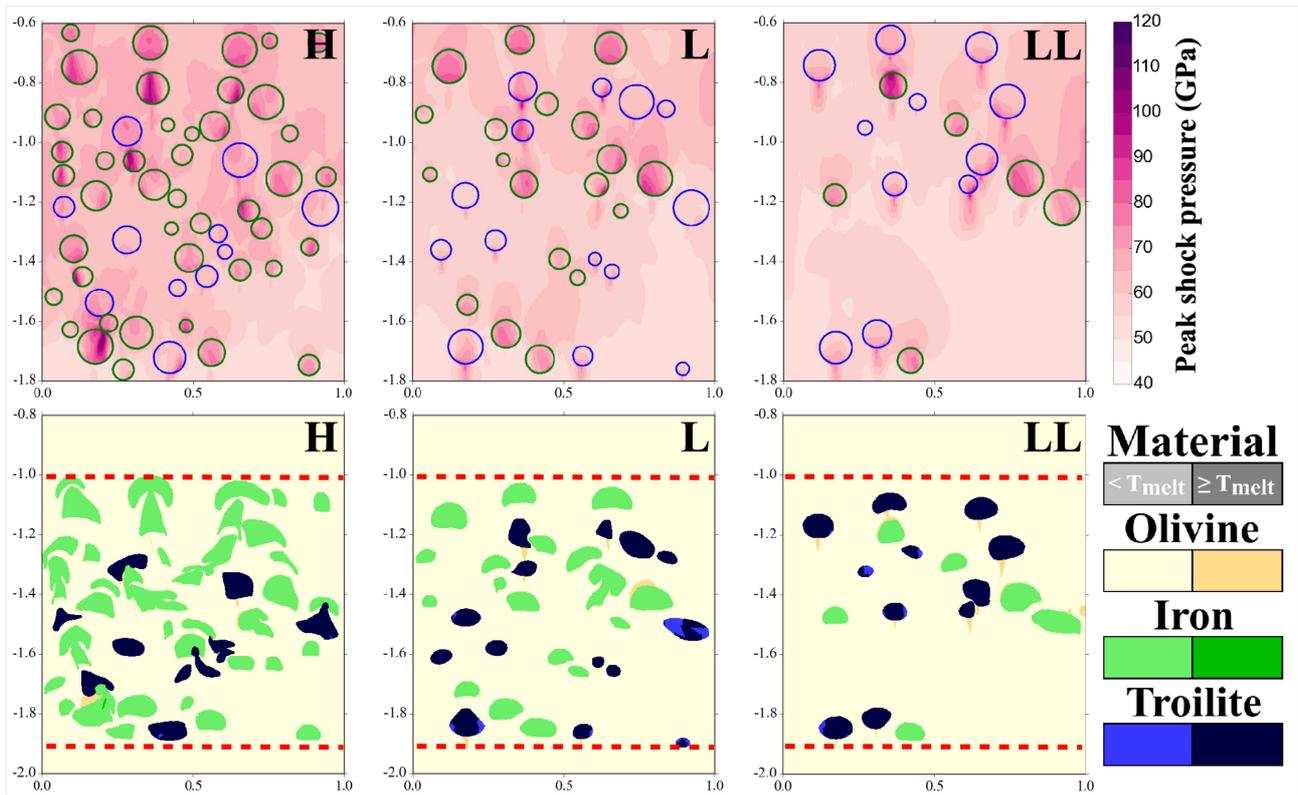

Fig. 9. 51 GPa nominal pressure ordinary chondrite meso-grain model results showing the sample plates (delineated by the red dashed lines at release) and particles (iron: green, troilite: blue in peak shock pressure panels). The peak shock pressures are shown in a non-compressed sample plate. All panels are at the same nominal pressure but the average generated peak shock pressures are different due to the strong reflections between olivine and iron/troilite grains (H-OC: 64 GPa, L-OC: 60 GPa, LL-OC: 58 GPa). Panels graduations are in mm. Each ordinary chondrite type is represented by its abbreviation. Color version of the figure is available in the electronic version of the manuscript.

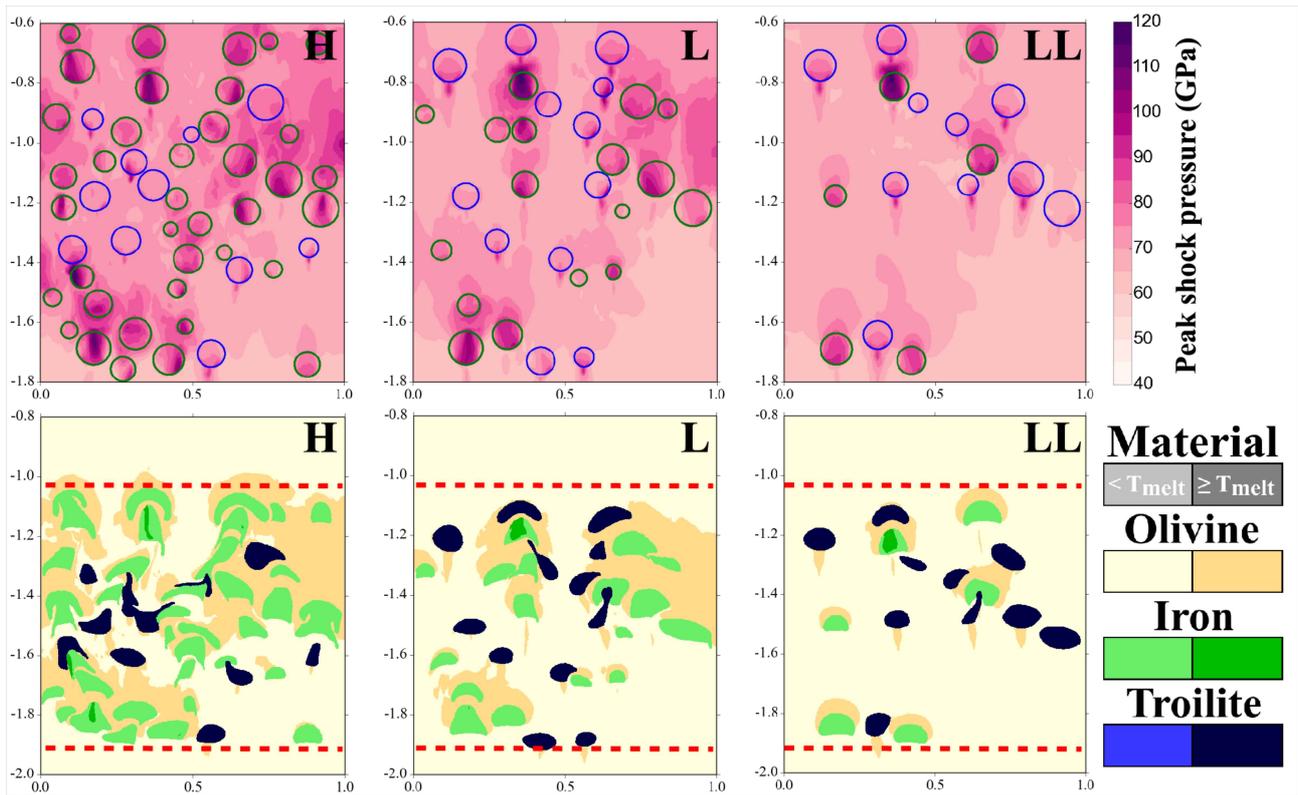

Fig. 10. 61 GPa nominal pressure ordinary chondrite meso-grain model results showing the sample plates (delineated by the red dashed lines at release) and particles (iron: green, troilite: blue in peak shock pressure panels). The peak shock pressures are shown in a non-compressed sample plate. All panels are at the same nominal pressure but the average generated peak shock pressures are different due to the strong reflections between olivine and iron/troilite grains (H-OC: 75 GPa, L-OC: 71 GPa, LL-OC: 68 GPa). Panels graduations are in mm. Each ordinary chondrite type is represented by its abbreviation. Color version of the figure is available in the electronic version of the manuscript.

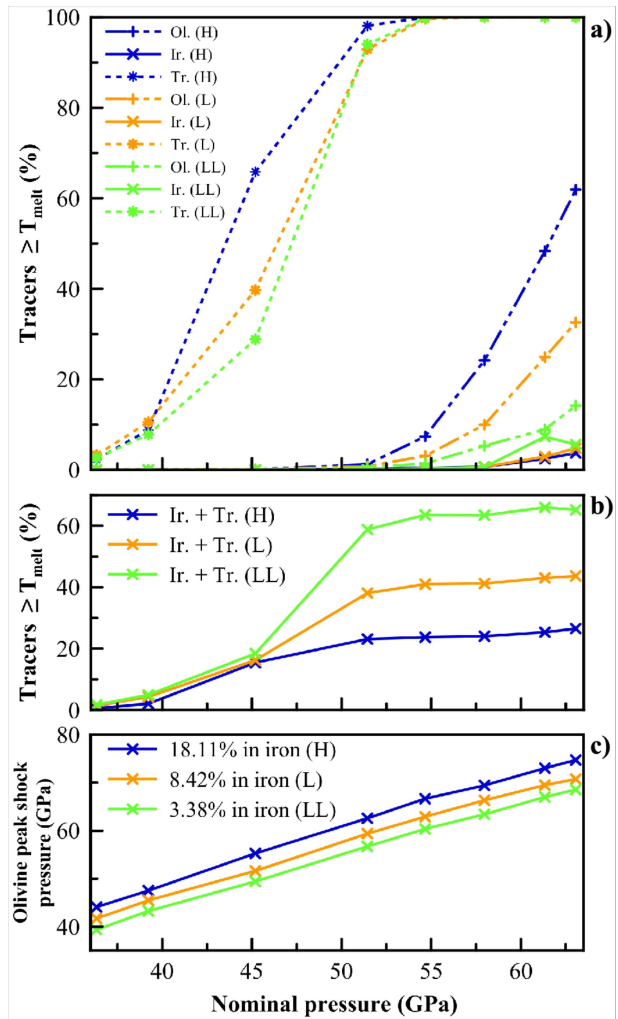

Fig. 11. Results of the ordinary chondrite mesoscale models using the meso-grains model: a) fractions of tracers $\geq T_{melt}$ for each material after pressure release in each ordinary chondrite type, b) combined troilite and iron melt-fractions to reflect the grain distribution and total shock-darkening melt phases and c) generated peak shock pressures in olivine in regard of the iron grains abundance. Each ordinary chondrite type is represented by its abbreviation and phases are Ir. - iron, Tr. - troilite, Ol. - olivine.

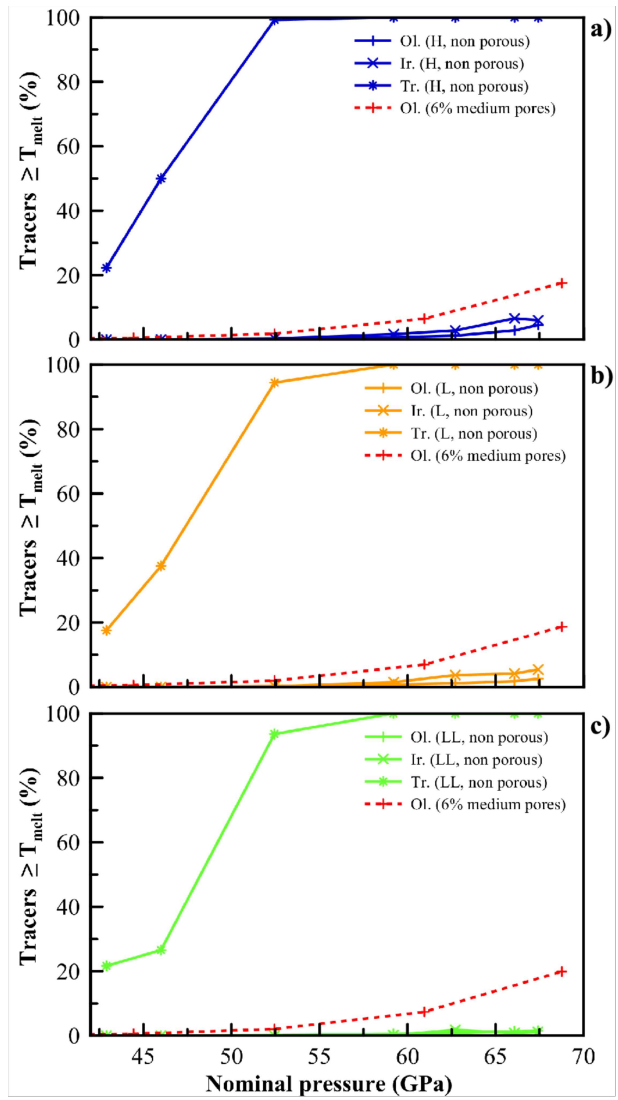

Fig. 12. Results of the ordinary chondrite mesoscale models using meso-grains/meso-pores model with fractions of tracers ≥ $T_{melt}$ of each material after pressure release. As a comparison the dotted lines are results for olivine in meso-pores model. Each ordinary chondrite type is represented by its abbreviation and phases are Ir. - iron, Tr. - troilite, Ol. - olivine.

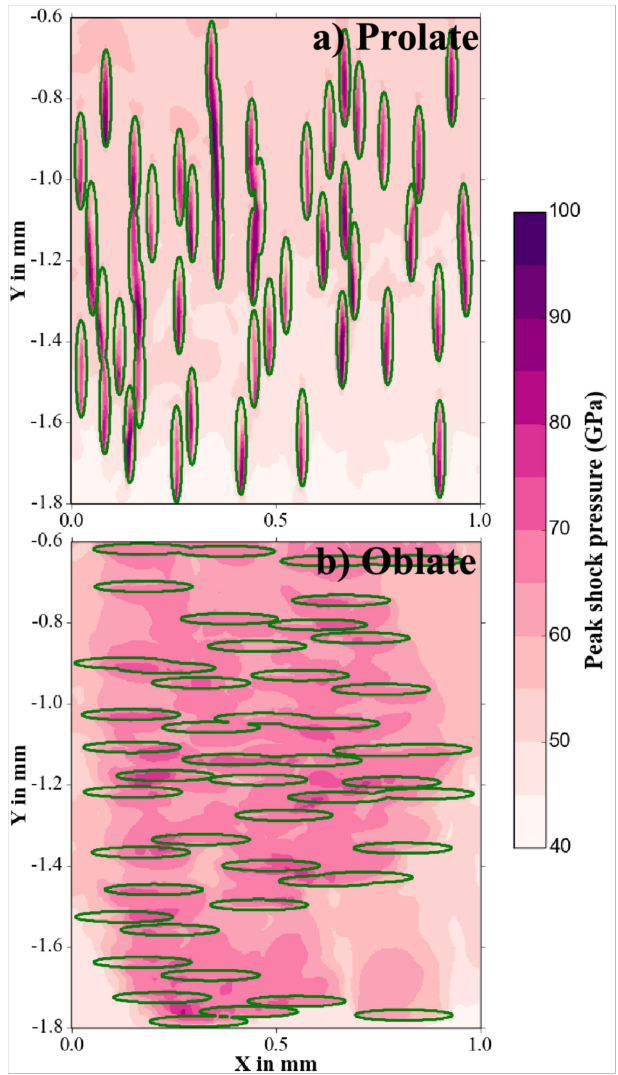

Fig. 13. Peak shock pressures obtained from a model with prolate and oblate iron grains in olivine matrix in the non-compressed sample plate. Nominal pressure is 45 GPa with average peak shock pressures of a) 54 GPa and b) 61 GPa. Iron grains are delineated by the green contours. Color version of the figure is available in the electronic version of the manuscript.

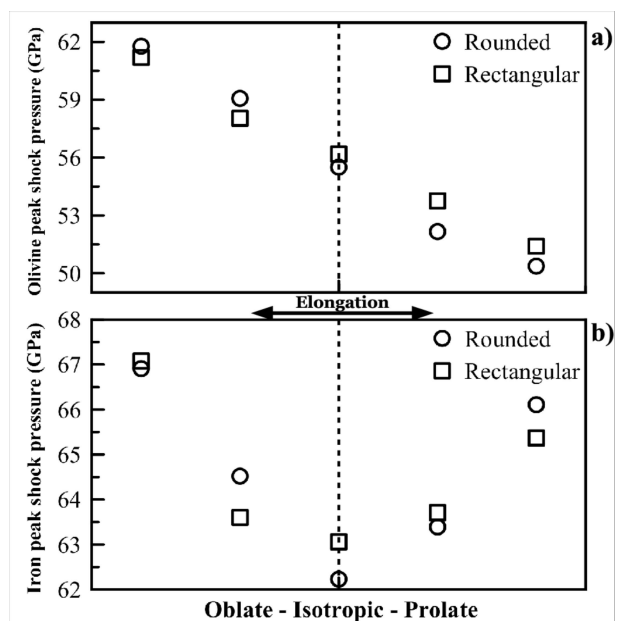

Fig. 14. a) Olivine and b) iron peak shock pressures in contrast to the shape and elongation of iron grains. Oblate grains are transversal to the shock wave and prolate grains are longitudinal to the shock wave.

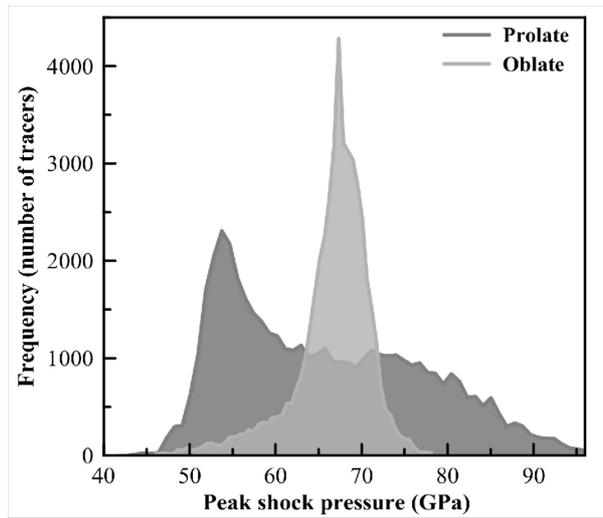

Fig. 15. Distribution of peak shock pressures in the most extreme oblate and prolate grain shape scenarios (Fig. 13, 14).

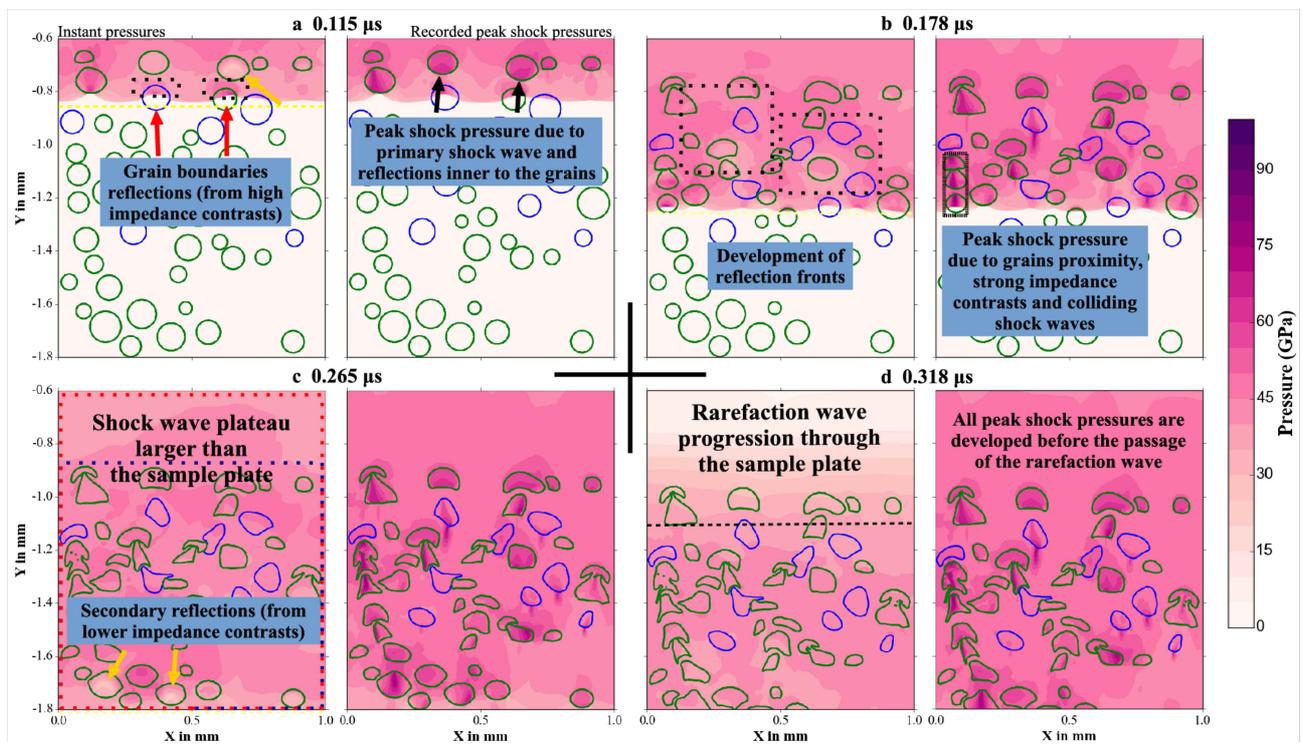

Fig. 16. Snapshots of a shock wave propagation in an H ordinary chondrite meso-grains model at 39.22 GPa nominal pressure. Each pair of graphics comprises the instant pressures (left) and the recorded peak shock pressures (right). The dashed yellow line is the shock front, arrows indicate areas of interest, black dotted boxes are two examples of the development of reflection fronts over time, the blue dotted box is the sample plate, the red dotted box is the shock wave plateau, and the dashed black line is the rarefaction wave front. Snapshots times are shown and the sample plate is also compressed and displaced in the direction of the shock front. Iron and troilite grains are delineated by green and blue lines respectively. Color version of the figure is available in the electronic version of the manuscript.

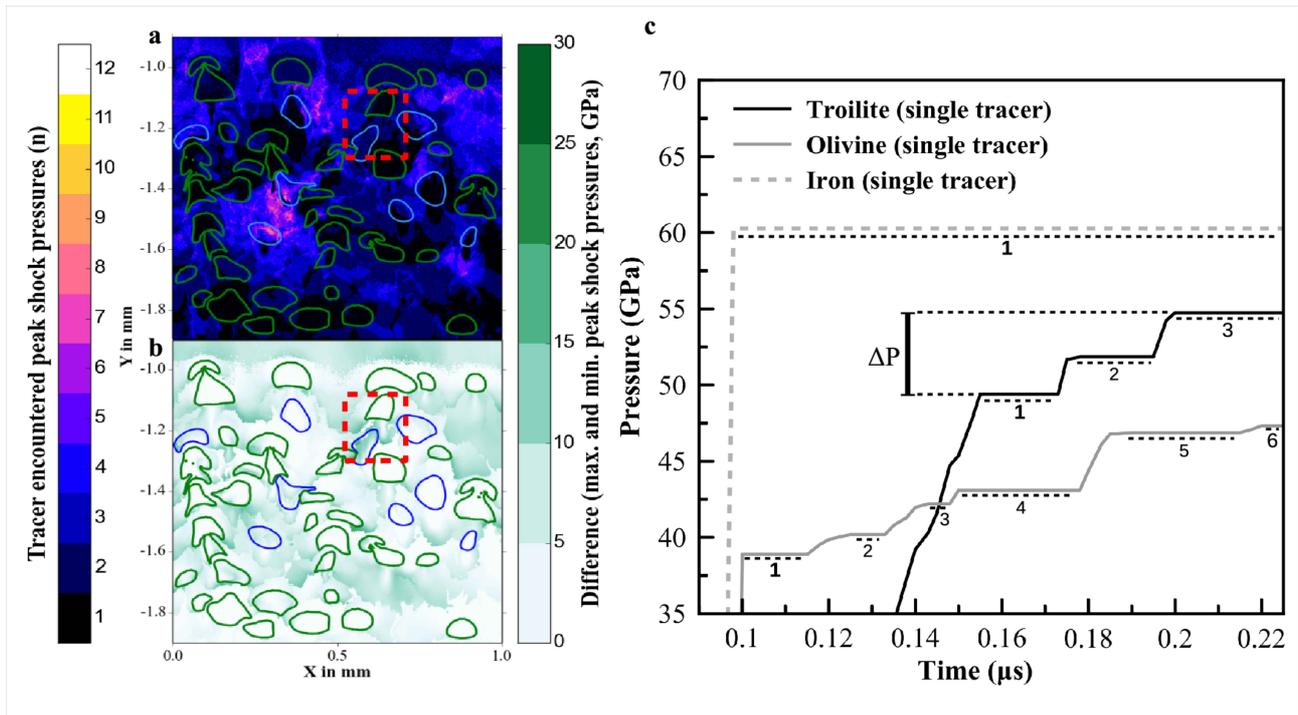

Fig. 17. Details and complexity of the peak shock pressures ramping in the sample plate after release (H ordinary chondrite meso-grains model at 39.22 GPa nominal pressure): a) peak shock pressures plateaus frequencies attained by each material unit tracer, b) differences between the lowest and highest peak pressure plateaus (Δp), c) examples of tracers recorded peak shock pressures over time. The fine dashed lines in c) represent the peak pressure plateaus. The last plateau of each tracer is the final recorded peak shock pressure used in the post-shock temperatures assessment. A peak shock pressure plateau shows a minimum of three occurrences of same value over time (0.005 μs total time lapse). It does not include reflections with lower pressures as seen in Fig. 16 (orange arrows). The dashed red boxes in a) and b) indicate an area of interest showing no specific dependence between a) and b). Color version of the figure is available in the electronic version of the manuscript.

# Supporting information document

Shock-darkening in ordinary chondrites: determination of the pressure-temperature conditions by shock physics mesoscale modeling

J. Moreau, T. Kohout and K. Wünnemann

**Post-shock temperatures calculations**

Post-shock temperatures are based on the assumption that the internal energy that remains in the material after the release from shock pressure is given by the difference between the *PdV* work for a given shock state and the integral under the adiabatic release path in *P-V* space, where *P* is pressure and *V* is specific volume. For simplicity the release path can be approximated by the Hugoniot curve (Raikes and Ahrens 1979) that is parameterized by a linear or piecewise linear relationship in the $u_p$-$U$ space, where $u_p$ is the particle velocity and *U* the shock wave velocity. The post-shock temperature can then be calculated from the post-shock internal energy with the assumption of a initial heat capacity at the initial temperature of the system (293K). Whilst ANEOS allows for more accurate temperature estimates if the models were run until the complete release state (release state method), the Tillotson EoS does not provide reliable temperature estimates. For this reason and for consistency among the processing of different materials with either Tillotson EoS or ANEOS we used the peak shock pressure method to estimate post-shock temperatures in all models in this study.

To do this we used the peak shock pressures recorded in tracers. We considered up to two phases transitions. Following the approach of Artemieva and Ivanov (2004) and Fritz et al. (2005) we list the employed parameters and equations below:

*General parameters:*

$\Delta T$: post-shock temperature (K)

$U$: shock wave velocity (m/s)

$P$: peak shock pressure (Pa) – recorded peak shock pressure in Lagrangian tracer

$P_0$: initial pressure (Pa) – 0 GPa

$c_p$: heat capacity (J/kgK) at room temperature

$\rho$: density (kg/m³) at $P_0$

$E_{shock}$: energy load at $P$

$H_f$: heat of fusion (J/kg)

$\alpha$: melt fraction (0-1)

*Material without LPP-HPP transition (low-pressure phase – high-pressure phase)*

$u_p$: particle velocity (m/s) at $P$

$C$: origin on y axis (m/s) in one phase material line $U = C + S.u_p$

$S$: slope in one phase material line $U = C + S.u_p$

$E_r$: energy release at $P$

*Material with LPP-HPP transition*

$P_{lim}$: pressure at phase change (Pa)

$u_{p\text{-}pmax1}$: particle velocity (m/s) in LPP at $P$ or $P_{lim}$

$u_{p\text{-}pmax2}$: particle velocity (m/s) in HPP at $P$

$C_{p1}$: origin on y axis (m/s) in LPP $U = C + S.u_p$

$C_{p2}$: origin on y axis (m/s) in HPP line $U = C + S.u_p$

$S_{p1}$: slope of LPP line $U = C + S.u_p$

$S_{p2}$: slope of HPP line $U = C + S.u_p$

$E_{r\text{-}p1}$: energy release in LPP at $P$ or $P_{lim}$

$E_{r\text{-}p2}$: energy release in HPP at $P$

*Calculation of post-shock temperature in one phase materials*

The calculation is based on the linear relationship between the particle velocity $u_p$ and propagation velocity of the shock front $U$ (1):

$$U = C + S.u_p \quad (1)$$

where the parameters $C$ and $S$ are determined from Hugoniot data generated by the EoS (see Fig. S1 for an example in iron and the corresponding chapter)

First, the particle velocity is calculated via (2):

$$u_p = -0{,}5\frac{C}{S} + \sqrt{0{,}25\frac{C^2}{S^2} + \frac{(P-P_0)}{\rho_0.S}} \quad (2)$$

The energy removed after the shock event (adiabatic decompression) is is given by (3):

$$E_R = \frac{C}{S}\left(u_p + \frac{C}{S}.\ln\left(\frac{S}{U}\right)\right) \quad (3)$$

where $U$ is obtained via (1) at $u_p$ from (2).

The energy at a given shock state (for a given $u_p$) is calculated by (4):

$$E_{shock} = \frac{u_p^2}{2} \quad (4)$$

Finally, $\Delta T$ is calculated via (5) with (3) and (4):

$$\Delta T = \frac{(E_{shock} - E_R)}{c_p} \quad (5)$$

As we used room temperature as a reference, 293 K is added to obtain the final temperature. The $\Delta T$ given by equation (5) is only reliable under the melting point of a material. Over the melting point, calculations become much more inaccurate as we do not consider the heat of fusion that is used to assess partial melting. Thus in our work we considered the melt fraction as the ratio between the number of tracers $\geq$ $T_{melt}$ and the total amount of tracers in a material (%). Also we did not consider varying heat capacity with temperature (see Main Document for more explanations).

If we include heat of fusion in equation (5), we can assess partial melting obtained as follows:

$$\alpha = \frac{E_{shock} - E_R - c_p(T_{melt} - T_0)}{H_f} \quad (6),$$

assuming that the upper term of equation (6), the energy remaining after reaching the melting point, is positive and not exceeding the heat of fusion.

*Calculation of post-shock temperature in materials with LPP-HPP transition below $P_{lim}$*

Post-shock temperatures are calculated from (1-5) and $P$ using $C_{p1}$ and $S_{p1}$ instead of $C$ and $S$.

*Calculation of post-shock temperature in materials with LPP-HPP transition over $P_{lim}$*

In this case we have to consider two energies at release. Two particles velocities are calculated. $u_{p\text{-}pmax1}$ is calculated via (2) using $C_{p1}$ and $S_{p1}$ instead of $C$ and $S$ and $P_{lim}$ instead of $P$. $u_{p\text{-}pmax2}$ is calculated via (2) using $C_{p2}$ and $S_{p2}$ instead of $C$ and $S$ and using $P$. Energy load $E_{shock}$ is calculated from (4) using $u_{p\text{-}pmax2}$. $E_{r\text{-}p1}$ and $E_{r\text{-}p2}$ are calculated using (7) and (8).

$$E_{R-p1} = \frac{C_{p1}}{S_{p1}} \left[ u_{p-pmax1} - \frac{C_{p1}}{S_{p1}} \cdot \ln\left(1 + \frac{S_{p1}}{C_{p1}} \cdot u_{p-pmax1}\right) \right] \quad (7)$$

$$E_{R-p2} = \frac{C_{p2}}{S_{p2}} \left[ u_{p-pmax2} - u_{p-pmax1} - \frac{C_{p2}}{S_{p2}} \cdot \ln\left( \frac{(C_{p2} + S_{p2} \cdot u_{p-pmax2})}{(C_{p2} + S_{p2} \cdot u_{p-pmax1})} \right) \right] \quad (8)$$

The post-shock temperature is then calculated using (9):

$$\Delta T = \frac{(E_{shock} - E_{R-p1} - E_{R-p2})}{c_p} \quad (9)$$

Equation (6) to assess partial melting can also be derived from equation (9).

*Determination of Cs and Ss parameters*

This process is explained in the Main Document (*Post-shock temperatures and melt assessment*). Fig. S1 shows an example of the linear relationship of $u_p$ and $U$ for iron. Dashed lines represent ANEOS data, the solid lines are the linear regressions with $C$ and $S$ parameters used for the methods described here to determine the post-shock temperature.

**Comparison with the release state method**

In order to assess the error we compared the release state and peak shock pressure methods (peak shock pressure). Advantages or disadvantages of these methods are:

1. Post-shock temperatures from the release state method can be obtained directly from the model and is accurate for ANEOS material. However this technique may suffer from numerical diffusion (true for Tillotson EoS). Localized "hotspots" may be smeared out over an increasingly larger area the longer the simulation lasts (in the Eulerian frame of reference, Collins et al. 2012).
2. In the peak shock pressure method, we only recorded peak shock pressures and so we did not need full release state. However due to reflections, the post-shock temperatures were overestimated if the peak pressure resulted from several superimposing shock waves. Thus it did not represent the pressure that were achieved by a single shock pulse. For an accurate treatment of such a case the intermediate shock stages should be treated separately to estimate the correct *PdV* work the material experienced to reach a certain peak shock pressure (piecewise integration).

3. In addition, the release state method requires to run the models until the material is completed unloaded, which is numerically much more expensive than running the models until the maximum shock pressures are achieved (peak shock pressure method).

As an example we simulated the propagation of a shock wave with nominal pressures of ~35, 45 and 55 GPa through a rounded iron (ANEOS) and troilite (Tillotson) grain embedded in an olivine matrix (ANEOS). We compared post-shock temperatures from the release state and peak shock pressure methods. In the release state method, we recorded temperatures at a threshold of 100 Pa to minimize numerical diffusion. We used the same setup as in our study and results are (Fig. S2, S3 and Table S1):

1. Iron post-shock temperatures showed some differences between the two methods but for the pressure range under consideration in this study the two methods correlated reasonably well. The differences resulted from simplified assumptions in the peak shock pressure method (constant heat capacity and approximation of the release adiabate by the Hugoniot curve). Furthermore, as seen in Fig. 17, we observed no reflections happening in the iron grain because the impedance contrast is strong with olivine.

2. Testing for troilite, we observed that models running to the very end were subject to strong numerical diffusion to the initial temperature. This made post-shock temperatures recording rather difficult and inaccurate (see standard deviation from Fig. S3). Thus it especially proved mandatory to use the peak shock pressure method for troilite in that case. No reflections occurred in troilite either.

3. Strong reflections occurred in olivine. Because we used peak shock pressure method for all materials in our study, post-shock temperatures were overestimated in olivine.

Note, physical heat diffusion is not taken into consideration in our simulations and cannot be approximated by numerical diffusion by no means.

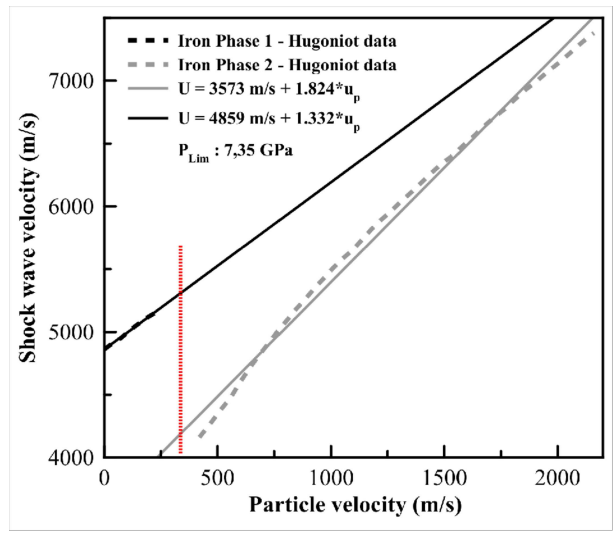

Fig. S1. Linear regressions on Hugoniot data in particle and shock wave velocity fields for iron phases (ANEOS). Red is the transition region. Each of these line parameters and transition pressure are used in post-shock temperatures calculation.

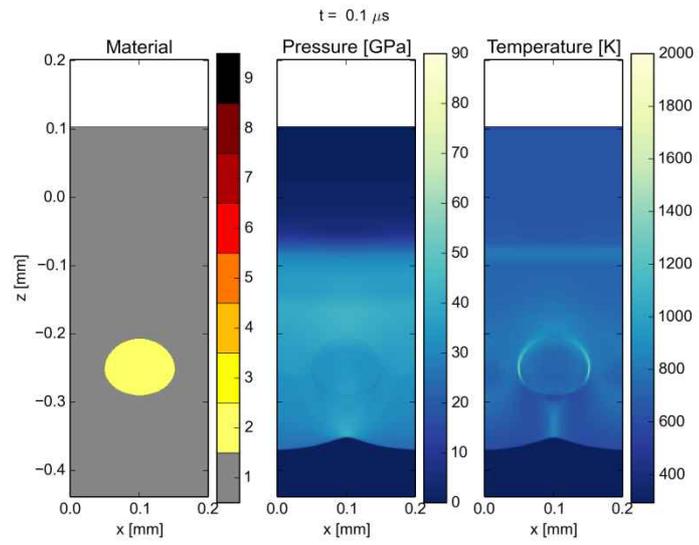

Fig. S2. Snapshots of a model using a single rounded grain of iron in olivine material at time step 0.1 microsecond. We observe some boundary heating of iron/olivine in this model. The grain is about 200 cells in diameter.

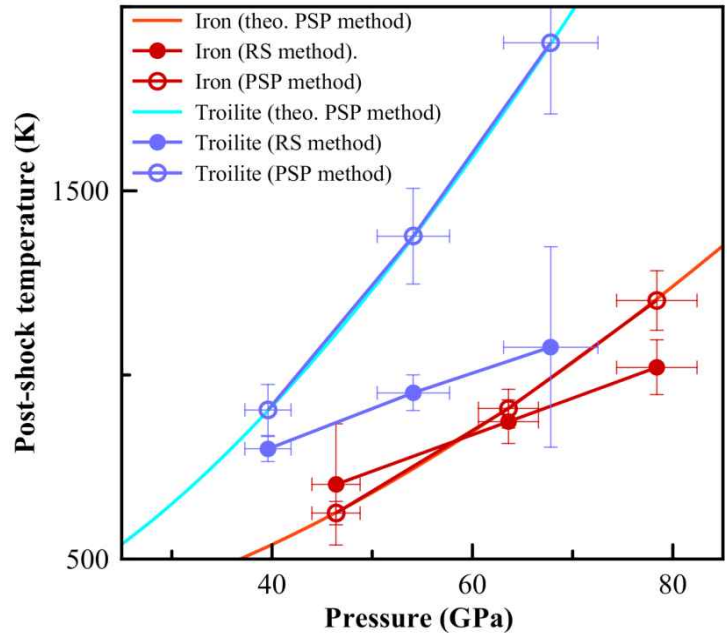

Fig. S3. Compilation of results on single rounded grains of iron or troilite showing release state (RS) and peak shock pressure (PSP) method post-shock temperatures. The strong deviation on troilite data is due to the Tillotson EoS definition unable to compute correct release temperatures.

Table S1. Release state (RS) and peak shock pressure (PSP) method post-shock temperatures.

| Description | Peak shock pressure (GPa) | std. dev. | Peak temperature (K) | std. dev. | RS method post-shock temperature (K) | std. dev. | PSP method post-shock temperature (K) | std. dev. |
|---|---|---|---|---|---|---|---|---|
| Iron rounded gr. | 46.4 | 2.4 | 802.8 | 88.7 | 702.7 | 164.9 | 625.0 | 32.2 |
| Iron rounded gr. | 63.6 | 3.0 | 1068.6 | 102.2 | 873.2 | 59.3 | 908.5 | 53.2 |
| Iron rounded gr. | 78.4 | 4.0 | 1323.1 | 104.9 | 1020.6 | 74.2 | 1202.3 | 80.8 |
| Troilite round. gr. | 39.6 | 2.3 | 1043.9 | 49.5 | 799.1 | 33.6 | 905.0 | 68.9 |
| Troilite round. gr. | 54.1 | 3.6 | 1502.2 | 82.3 | 951.6 | 48.3 | 1377.4 | 129.9 |
| Troilite round. gr. | 67.8 | 4.7 | 2012.5 | 131.1 | 1075.5 | 272.3 | 1902.1 | 192.7 |

**Supplementary data (referred in manuscript)**

Table S2. General material parameters.

| | Olivine $Fo_{75}Fa_{25}$ | Enstatite | Troilite | Iron | Olivine $Fo_{90}Fa_{10}$ |
|---|---|---|---|---|---|
| ***Tillotson parameters*** | | | | *(ANEOS)* | *(ANEOS)* |
| Density $\rho$ (g/cm³) | $3.491*10^3$ | $3.34*10^{3\,(1)}$ | $4.611*10^{3\,(4)}$ ($4.7*10^3$) | $7.840*10^3$ | $3.314*10^3$ |
| Bulk modulus $A$ (Pa) | $1.308*10^{11}$ | $1.058*10^{11}$ | $0.45*10^{11\,(4)}$ ($0.8*10^{11}$) | - | - |
| Tillotson parameter $B$ (Pa) | $4.9*10^{10}$ | $4.9*10^{10}$ | $4.0*10^{10}$ | - | - |
| Internal energy $E_0$ (J/kg) | $5.5*10^8$ or $10.0193*10^6$ | $5.5*10^8$ or $10.0193*10^6$ | $14.343*10^6$ | - | - |
| Tillotson parameter $a$ | 0.5 | 0.5 | 0.5 | - | - |
| Tillotson parameter $b$ | 1.4 | 1.4 | 1.4 | - | - |
| Tillotson parameter $\alpha$ | 5 | 5 | 5 | - | - |
| Tillotson parameter $\beta$ | 5 | 5 | 5 | - | - |
| Incipient vaporization Internal energy $E_{IV}$ (J/kg) | $4.5*10^6$ | $4.5*10^6$ | $3*10^6$ | - | - |
| Complete vaporization Internal energy $E_{CV}$ (J/kg) | $1.45*10^7$ | $1.45*10^7$ | $5*10^6$ | - | - |
| ***Thermal parameters*** | | | | | |
| Heat capacity $c_p$ (J/kg/K) | 816 | 796 [1] | 619.23 | 449 | 814 [6] |
| Melting temperature $T_{melt}$ (K) | 2000.75 | 1832 | 1463 | 1825 [5] | 2049 [6] |
| Simon's parameter $a$ (GPa) | 11.3429 | 2.85 [2] | 3 | 57.753 [5] | 11.175 |
| Simon's parameter $c$ | 2.9889 | 5.01 [2] | 4 | 1.529 [5] | 3.1713 |
| ***Strength parameters*** | | | | | |
| Yield strength $Y$ (Pa) | $1.5*10^9$ | $1.5*10^9$ | Hydrodyn. | Hydrodyn. | $1.5*10^9$ |
| Poisson ratio $\upsilon$ | 0.264 | 0.224 [3] | - | - | 0.257 [6,7] |

[1] for hypersthene (Enstatite $En_{85-75}$)
[2] Boyd et al. (1964)
[3] calculated for pyroxene $En_{77}Fs_{23}$ between enstatite and ferrosilite
[4] density of pyrrhotite (slightly lower that troilite, $4.7*10^3$ g/cm³) and bulk modulus adapted to fit the Hugoniots data.
[5] Zhang et al. (2015)
[6] values given for H-chondrites.
[7] Christensen (1996)

Table S3. Mesoscale model averaged results with porous olivine corresponding to H-chondrite composition.

| Nominal pressure (GPa) | 36.3 | 39.2 | 45.2 | 51.4 | 54.7 | 58.0 | 61.3 | 63.1 |
|---|---|---|---|---|---|---|---|---|
| (Bulk peak shock pressure) | 45.4 | 48.9 | 56.9 | 64.6 | 68.7 | 71.5 | 75.1 | 76.9 |
| **Olivine** | | | | | | | | |
| Tracers ≥ $T_{melt}$ (%) | 0 | 0.02 | 0.1 | 1.25 | 7.4 | 24.21 | 48.35 | 61.9 |
| Peak shock pressure (GPa) | 44.1 | 47.5 | 55.3 | 62.6 | 66.7 | 69.4 | 73.0 | 74.7 |
| **Iron** | | | | | | | | |
| Fraction (%) | 18.12 | 18.12 | 18.04 | 18.1 | 18.14 | 18.12 | 18.11 | 18.09 |
| Tracers ≥ $T_{melt}$ (%) | 0 | 0 | 0 | 0.17 | 0.43 | 0.68 | 2.49 | 3.74 |
| Peak shock pressure (GPa) | 50.3 | 54.5 | 63.1 | 72.0 | 76.3 | 79.3 | 83.4 | 84.9 |
| **Troilite** | | | | | | | | |
| Fraction (%) | 5.53 | 5.58 | 5.53 | 5.54 | 5.54 | 5.57 | 5.54 | 5.61 |
| Tracers ≥ $T_{melt}$ (%) | 2.36 | 8.81 | 65.82 | 98.1 | 99.99 | 100 | 100 | 100 |
| Peak shock pressure (GPa) | 47.7 | 50.4 | 59.0 | 67.5 | 71.4 | 73.8 | 76.9 | 80.1 |
| **Total iron/troilite ≥ $T_{melt}$ (%)** | 0.55 | 2.07 | 15.44 | 23.12 | 23.72 | 24.03 | 25.33 | 26.53 |
| **Total melt (%)** | 0.13 | 0.51 | 3.72 | 6.42 | 11.27 | 24.17 | 42.91 | 53.52 |

Table S4. Mesoscale model averaged results with non-porous olivine corresponding to H-chondrite composition.

| Nominal pressure (GPa) | 42.9 | 46.0 | 52.5 | 59.2 | 62.7 | 66.1 | 67.4 |
|---|---|---|---|---|---|---|---|
| (Bulk peak shock pressure) | 51.6 | 55.2 | 63.5 | 70.3 | 73.2 | 76.4 | 77.6 |
| **Olivine** | | | | | | | |
| Tracers ≥ $T_{melt}$ (%) | 0.02 | 0.04 | 0.26 | 0.75 | 1.23 | 2.87 | 4.58 |
| Peak shock pressure (GPa) | 49.9 | 53.5 | 61.6 | 68.2 | 71.0 | 74.2 | 75.7 |
| **Iron** | | | | | | | |
| Fraction (%) | 18.11 | 18.14 | 18.05 | 18.11 | 18.06 | 18.07 | 18.16 |
| Tracers ≥ $T_{melt}$ (%) | 0 | 0.02 | 0.34 | 1.7 | 2.86 | 6.6 | 6.03 |
| Peak shock pressure (GPa) | 57.6 | 61.6 | 70.3 | 78.2 | 81.4 | 85.3 | 85.4 |
| **Troilite** | | | | | | | |
| Fraction (%) | 5.64 | 5.6 | 5.57 | 5.55 | 5.58 | 5.51 | 5.65 |
| Tracers ≥ $T_{melt}$ (%) | 22.28 | 49.98 | 99.21 | 100 | 100 | 100 | 100 |
| Peak shock pressure (GPa) | 54.7 | 58.0 | 68.0 | 74.1 | 76.9 | 78.3 | 79.5 |
| **Total iron/troilite ≥ $T_{melt}$ (%)** | 5.29 | 11.81 | 23.66 | 24.76 | 25.79 | 28.43 | 27.97 |
| **Total melt (%)** | 1.27 | 2.83 | 5.79 | 6.43 | 7.04 | 8.90 | 10.09 |

Table S5. Mesoscale model averaged results with porous olivine corresponding to L-chondrite composition.

| Nominal pressure (GPa) | 36.3 | 39.2 | 45.2 | 51.4 | 54.7 | 58.0 | 61.3 | 63.1 |
|---|---|---|---|---|---|---|---|---|
| (Bulk peak shock pressure) | 42.6 | 46.5 | 52.8 | 60.7 | 64.3 | 67.8 | 71.0 | 72.3 |
| **Olivine** | | | | | | | | |
| Tracers ≥ $T_{melt}$ (%)Melt (%) | 0 | 0.02 | 0.11 | 0.8 | 3.11 | 9.98 | 24.88 | 32.6 |
| Peak shock pressure (GPa) | 41.7 | 45.5 | 51.6 | 59.4 | 62.9 | 66.3 | 69.5 | 70.7 |
| **Iron** | | | | | | | | |
| Fraction (%) | 8.45 | 8.43 | 8.41 | 8.39 | 8.36 | 8.45 | 8.39 | 8.44 |
| Tracers ≥ $T_{melt}$ (%) | 0 | 0 | 0 | 0 | 0.06 | 0.69 | 2.94 | 4.79 |
| Peak shock pressure (GPa) | 49.3 | 54.2 | 62.9 | 71.2 | 75.4 | 78.8 | 83.0 | 85.7 |
| **Troilite** | | | | | | | | |
| Fraction (%) | 5.85 | 5.81 | 5.83 | 5.83 | 5.83 | 5.82 | 5.9 | 5.82 |
| Tracers ≥ $T_{melt}$ (%) | 3.28 | 10.55 | 39.72 | 92.91 | 99.63 | 100 | 100 | 100 |
| Peak shock pressure (GPa) | 45.5 | 50.2 | 56.1 | 64.5 | 68.2 | 72.6 | 75.2 | 76.4 |
| **Total iron/troilite ≥ $T_{melt}$ (%)** | 1.34 | 4.30 | 16.26 | 38.09 | 40.97 | 41.19 | 43.01 | 43.65 |
| **Total melt (%)** | 0.19 | 0.63 | 2.41 | 6.10 | 8.48 | 14.43 | 27.47 | 34.18 |

Table S6. Mesoscale model averaged results with non-porous olivine corresponding to L-chondrite composition.

| Nominal pressure (GPa) | **42.9** | **46.0** | **52.5** | **59.2** | **62.7** | **66.1** | **67.4** |
|---|---|---|---|---|---|---|---|
| (Bulk peak shock pressure) | 48.7 | 52.2 | 59.7 | 67.5 | 70.7 | 72.9 | 73.8 |
| **Olivine** | | | | | | | |
| Tracers ≥ $T_{melt}$ (%) | 0.02 | 0.04 | 0.18 | 0.75 | 1.15 | 1.79 | 2.52 |
| Peak shock pressure (GPa) | 47.7 | 51.1 | 58.4 | 66.1 | 69.2 | 71.4 | 72.2 |
| **Iron** | | | | | | | |
| Fraction (%) | 8.39 | 8.39 | 8.39 | 8.44 | 8.47 | 8.37 | 8.42 |
| Tracers ≥ $T_{melt}$ (%) | 0 | 0.03 | 0.22 | 1.47 | 3.69 | 4.15 | 5.42 |
| Peak shock pressure (GPa) | 56.0 | 60.7 | 69.1 | 78.3 | 81.4 | 84.0 | 86.9 |
| **Troilite** | | | | | | | |
| Fraction (%) | 5.78 | 5.87 | 5.82 | 5.83 | 5.83 | 5.85 | 5.78 |
| Tracers ≥ $T_{melt}$ (%) | 17.54 | 37.55 | 94.43 | 100 | 100 | 100 | 100 |
| Peak shock pressure (GPa) | 52.9 | 56.3 | 64.3 | 73.5 | 76.7 | 79.5 | 77.7 |
| **Total iron/troilite ≥ $T_{melt}$ (%)** | 7.15 | 15.47 | 38.81 | 41.72 | 42.95 | 43.58 | 43.92 |
| **Total melt (%)** | 1.03 | 2.24 | 5.67 | 6.60 | 7.13 | 7.73 | 8.40 |

Table S7. Mesoscale model averaged results with porous olivine corresponding to LL-chondrite composition.

| Nominal pressure (GPa) | **36.3** | **39.2** | **45.2** | **51.4** | **54.7** | **58.0** | **61.3** | **63.1** |
|---|---|---|---|---|---|---|---|---|
| (Bulk peak shock pressure) | 40.0 | 43.9 | 50.2 | 57.7 | 61.3 | 64.4 | 68.2 | 69.7 |
| **Olivine** | | | | | | | | |
| Tracers ≥ $T_{melt}$ (%) | 0 | 0 | 0.04 | 0.56 | 1.42 | 5.34 | 8.95 | 14.19 |
| Peak shock pressure (GPa) | 39.4 | 43.3 | 49.4 | 56.7 | 60.4 | 63.4 | 67.0 | 68.6 |
| **Iron** | | | | | | | | |
| Fraction (%) | 3.32 | 3.38 | 3.31 | 3.48 | 3.35 | 3.39 | 3.39 | 3.41 |
| Tracers ≥ $T_{melt}$ (%) | 0 | 0 | 0 | 0 | 0.34 | 0.62 | 7.36 | 5.61 |
| Peak shock pressure (GPa) | 51.3 | 53.6 | 64.2 | 71.5 | 77.2 | 80.0 | 88.2 | 88.1 |
| **Troilite** | | | | | | | | |
| Fraction (%) | 5.89 | 5.88 | 5.84 | 5.81 | 5.83 | 5.82 | 5.84 | 5.83 |
| Tracers ≥ $T_{melt}$ (%) | 2.66 | 7.82 | 28.85 | 94.02 | 99.82 | 100 | 100 | 100 |
| Peak shock pressure (GPa) | 43.8 | 48.7 | 54.9 | 63.9 | 67.1 | 71.5 | 74.2 | 76.5 |
| **Total iron/troilite ≥ $T_{melt}$ (%)** | 1.70 | 4.97 | 18.41 | 58.80 | 63.52 | 63.42 | 65.98 | 65.17 |
| **Total melt (%)** | 0.16 | 0.46 | 1.72 | 5.97 | 7.12 | 10.69 | 14.21 | 18.90 |

Table S8. Mesoscale model averaged results with non-porous olivine corresponding to LL-chondrite composition.

| Nominal pressure (GPa) | **42.9** | **46.0** | **52.4** | **59.2** | **62.7** | **66.1** | **67.4** |
|---|---|---|---|---|---|---|---|
| (Bulk peak shock pressure) | 47.4 | 50.3 | 57.6 | 64.4 | 68.0 | 70.7 | 71.5 |
| **Olivine** | | | | | | | |
| Tracers ≥ $T_{melt}$ (%) | 0.03 | 0.05 | 0.23 | 0.61 | 0.85 | 1.4 | 1.56 |
| Peak shock pressure (GPa) | 46.7 | 49.5 | 56.8 | 63.4 | 67.0 | 69.7 | 70.4 |
| **Iron** | | | | | | | |
| Fraction (%) | 3.38 | 3.31 | 3.34 | 3.43 | 3.35 | 3.31 | 3.33 |
| Tracers ≥ $T_{melt}$ (%) | 0 | 0 | 0.02 | 0.2 | 1.81 | 0.84 | 1.18 |
| Peak shock pressure (GPa) | 54.7 | 62.9 | 68.3 | 79.4 | 82.8 | 83.9 | 85.8 |
| **Troilite** | | | | | | | |
| Fraction (%) | 5.88 | 5.9 | 5.87 | 5.92 | 5.86 | 5.79 | 5.9 |
| Tracers ≥ $T_{melt}$ (%) | 21.64 | 26.57 | 93.58 | 100 | 100 | 100 | 100 |
| Peak shock pressure (GPa) | 53.1 | 55.3 | 64.4 | 71.9 | 75.2 | 78.8 | 79.8 |
| **Total iron/troilite ≥ $T_{melt}$ (%)** | 13.74 | 17.02 | 59.65 | 63.39 | 64.28 | 63.93 | 64.35 |
| **Total melt (%)** | 1.30 | 1.61 | 5.70 | 6.48 | 6.69 | 7.09 | 7.36 |